\documentclass[11pt]{article}
\usepackage{amsmath, amssymb}
\usepackage[a4paper, margin=1in]{geometry}  
\usepackage{braket}
\usepackage[hidelinks]{hyperref}
\usepackage{natbib}
\usepackage{graphicx}
\usepackage[dvipsnames]{xcolor}
\usepackage{svg}
\usepackage{subcaption}
\usepackage{comment}

\begin{document}

\title{Dynamics of Quantum Analogs of Classical Impact Oscillators}

\author{Arnab Acharya\footnote{dr.arnab.acharya@gmail.com} ,
Titir Mukherjee\footnote{tm21rs050@iiserkol.ac.in} ,
Deepshikha Singh\footnote{ds21ms133@iiserkol.ac.in} , 
Soumitro Banerjee\footnote{Corresponding author, soumitro@iiserkol.ac.in\\
Department of Physical Sciences, Indian Institute of Science Education \& Research, Kolkata, India
}}




\date{}

\maketitle

\begin{abstract}
This paper investigates the dynamics of quantum analogs of classical impact oscillators to explore how complex nonlinear behaviors manifest in quantum systems. While classical impact oscillators exhibit chaos and bifurcations, quantum systems, governed by linear equations, appear to forbid such dynamics. Through simulations of unforced, forced, and dissipative quantum oscillators, we uncover quasiperiodicity, strange nonchaotic dynamics, and even chaos in the presence of dissipation. Using entropy time series, Fourier spectra, OTOCs, Lyapunov analysis, and the 0–1 test, we demonstrate that quantum systems can exhibit rich dynamical signatures analogous to classical nonlinear systems, bridging quantum mechanics and chaos theory.
\end{abstract}



\section{Introduction}

It is known that classical nonlinear systems can exhibit an array of complex dynamical behaviors, including limit cycles, quasiperiodicity, strange nonchaotic orbits, and chaos. In contrast, it is believed that quantum systems cannot exhibit such dynamics because of their linear nature. This poses a fascinating puzzle: If the classical world emerges from the quantum world, how can complex dynamical behaviors emerge? After all, classical systems are built on quantum systems at a fundamental level! To probe this issue, we ask: What dynamical phenomena can a quantum system exhibit?

The field of quantum chaos seeks to answer this question by identifying the quantum signatures (mainly in terms of the distribution of energy levels) in systems whose classical counterparts are chaotic \cite{Stockmann_1999,haake1991quantum,cvitanovic2001classical}. For example, \cite{Bohigas1984} conjectured that the energy level statistics of a classically chaotic quantum system follow the predictions of random matrix theory (RMT). While trying to explain the spectra of complex nuclei, Eugene Wigner came up with the idea of representing the Hamiltonian with a Hermitian matrix whose elements are drawn randomly from a normal distribution. Subsequently, this approach has led to successful predictions regarding the statistics of eigenvalues of complex quantum systems. 

It has been found that in systems whose classical limits are not chaotic, the eigenvalues are uncorrelated, and their spacings follow the Poisson distribution ($P(s) \sim e^{-s}$, where $s$ is the normalized level spacing). In systems with a chaotic classical limit, the spacings of the energy levels follow either the Gaussian Unitary Ensemble (GUE), the Gaussian Orthogonal Ensemble (GOE), or the Gaussian Symplectic Ensemble (GSE)~\cite{Mehta2004} depending on the symmetries of the system. When the system has no symmetries, the Hamiltonian is a Hermitian matrix with complex entries, and the GUE statistics are followed. If the system has time-reversal symmetry, the matrices are real-valued symmetric, and GOE is followed. For time-reversal symmetric systems with rotational/spin symmetries, the matrices have quaternionic entries, and GSE is followed. Each ensemble has characteristic level spacing statistics: for GOE, $P(s) = \frac{\pi s}{2} \exp\left(-\frac{\pi s^2}{4}\right)$; for GUE, $P(s) = \frac{32 s^2}{\pi^2} \exp\left(-\frac{4 s^2}{\pi}\right)$; and for GSE, $P(s) = \frac{2^{18}}{3^6 \pi^3} s^4 \exp\left(-\frac{64 s^2}{9\pi}\right)$.

Additionally, there is a deep connection between the classical unstable periodic orbits of chaotic systems and the quantum density of states (energy spectrum in the case of bound states) ~\cite{gutzwiller2013chaos}. Another prominent connection is the presence of anomalously high probability densities along certain unstable classical periodic orbits, called quantum scars ~\cite{heller1984bound, kaplan1999measuring}. 

In our work, we adopt a different route. 
The `state' of a quantum system is represented by the wavefunction $\Psi(x,t)$, a complex-valued function of space and time. The system dynamics would be given by the evolution of the wavefunction following the Schr\"odinger equation. We try to infer the qualitative character of the system dynamics from the oscillation of the wavefunction.

Any study requires a model system. For our purpose, it needs to be a simple classical system whose quantum analog can be easily constructed. We chose the impact oscillator as our model because it is a very simple mechanical system that exhibits rich dynamics due to grazing-induced bifurcations \cite{nordmark1991non,chin1994grazing,bernardo2008piecewise}.

In order to apply the established methods of nonlinear dynamics, we need to convert the dynamics of
the complex-valued wavefunction into a real-valued time series. This can be achieved by using the expectation values of the observables \cite{shankar2012principles}, the autocorrelation function \cite{nauenberg1990autocorrelation}, the $L_1$-norm \cite{Horn_Johnson_1985}, and quantum entropies \cite{chehade2019quantum}. After trying various possibilities, we decided to use the entropy to generate a real-valued time series because of its suitability to distinguish qualitatively different types of dynamics. 

In classical dynamics, Takens' embedding theorem allows us to reconstruct the phase space of a system from a single time series. Although a direct quantum analog is not straightforward, we can visualize quantum dynamics in a phase-space-like representation using quasiprobability distributions.
The most famous of these is the Wigner function, \(W(x, p)\), which provides a `quantum phase space portrait' of the system. Although it is not a true probability distribution (it can take on negative values, a clear sign of quantum interference), its evolution can reveal important dynamical features. Studies have shown that the structure of the Wigner function can reflect the underlying classical dynamics, with quantum interference effects becoming more pronounced in chaotic systems and it can also exhibit `scars' \citep{heller1984bound}.

The Lyapunov exponent (LE), which quantifies the rate of exponential divergence of classical trajectories, has a quantum counterpart in the form of an out-of-time-order correlator (OTOC). The OTOC, defined as
\begin{equation}
    C(t) \equiv -\langle [\widehat{W}(t), \widehat{V}(0)]^2 \rangle , \label{otoc1}
\end{equation}
measures how a small local perturbation affects a later measurement. 
Here, \( \widehat{V}(0) \) and \( \widehat{W}(t) \) are two Hermitian operators in the Heisenberg picture at times 0 and $t$, and the brackets \(\langle \cdot \rangle\) represent either a quantum-mechanical expectation value or a thermal average, depending on the context \cite{hashimoto2017out}. In a system with periodic dynamics, the OTOC varies periodically or quasiperiodically with time \cite{hashimoto2017out,li2023out}. In quasiperiodically driven systems, the OTOC exhibits power-law growth at early times, as observed in the Aubry–Andr\'e model \cite{riddell2020out}. In systems where OTOC grows exponentially in time as \(C(t) \sim e^{\lambda_L t}\) before saturating, the dynamics is typically considered chaotic \cite{akutagawa2020out}.

Another method of distinguishing different types of dynamics is to analyze the frequency spectrum. A periodic orbit is expected to have a fundamental component and its harmonics; a quasiperiodic orbit is supposed to have more than one fundamental frequency and combinations of their harmonics; a strange nonchaotic orbit has dense but discrete peaks in the spectrum \cite{prasad2001strange}; and a chaotic orbit has a spread spectrum---a continuum of frequency components \cite{huberman1981powerspectra}.

The 0-1 test \cite{gottwald2004new} has been proposed as a reliable method of distinguishing between chaotic and nonchaotic time series. It takes
a sampled data set as input and outputs a single value, $K$, between 0 and
1, with 0 indicating periodicity/quasi-periodicity and 1 indicating chaos. No
prior information about the system is required to apply the test. This test proved to be particularly useful in our case, as the entropy time series is all we have to diagnose the qualitative character of the system.

Bifurcation diagrams are a staple of nonlinear dynamics, illustrating how a system's behavior changes as a parameter is varied. Although the linearity of the Schr\"odinger equation prevents true bifurcations in the classical sense, analogous phenomena can be observed in quantum systems through parameter-dependent changes in their properties.
By systematically changing a parameter in a quantum system (like the strength of a driving field), we can observe abrupt changes in the system's properties, such as its energy level structure or the morphology of its wavefunctions. These transitions, often termed quantum bifurcations, can be visualized in diagrams that bear a striking resemblance to their classical counterparts. For example, studies of molecular systems have shown how nonadiabatic couplings can lead to `bifurcation' and merging of quantum wave packets, inducing complex dynamics \citep{Takami2024}.


Earlier studies on the classical impact oscillator considered a friction or dissipative element and demonstrated an abrupt transition to chaos at grazing \cite{nordmark,bernardo2008piecewise,banerjee2009invisible}. 
In quantum mechanics, dissipation is typically addressed by modeling the system as open, i.e., interacting with an external environment or bath. This bath is often represented as a collection of harmonic oscillators. Such a framework enables researchers to explore how the dynamical properties evolve in the presence of decoherence and energy loss.

In recent years, the study of quantum chaos in dissipative settings has shown that dissipation, while suppressing coherence, can also sustain chaotic signatures for longer durations~\cite{chaos_dissipation2025}, as demonstrated in many-body systems where chaos is linked to decoherence, entanglement, and thermalization~\cite{manybody_theory2025, manybody_experiment2024}. Classical chaotic systems such as the kicked top~\cite{kicked_top1, kicked_top2}, Duffing oscillator~\cite{duffing1, duffing2}, and the Morse oscillator~\cite{morse_oscillator} have long served as benchmarks for exploring the correspondence between classical and quantum in dissipation. However, in comparison, the soft impact oscillator---despite being a canonical piecewise-smooth system in the classical domain---has received relatively little attention in the quantum dissipative regime, motivating the present study.

The standard framework for studying open quantum systems is the Lindblad master equation \cite{manzano2020short,nafari2022enhancing}, which is based on the construction of suitable jump operators to represent environmental effects. While this works well for systems with certain symmetries, it becomes impractical for cases like the quantum impact oscillator, where the lack of symmetry leads to an infinite and computationally expensive expansion.

To overcome these challenges, we adopt the quantum Langevin equation (QLE) formalism, which offers a tractable alternative for modeling dissipative dynamics in open quantum systems. We consider a system coupled to a bath of harmonic oscillators and, following standard procedures \cite{Ford1988, agarwal1971brownian}, transition to the Heisenberg picture. By deriving the equations of motion for both system and bath operators and subsequently eliminating the bath degrees of freedom, we obtain an effective description of the dynamics of the system.

We employ all of the above methods to analyze the dynamics of the quantum system. The paper is organized as follows. In Section~\ref{impact}, we introduce the classical impact oscillator and its quantum analog. In Section~\ref{unforced} we analyze the dynamics of the quantum impact oscillator and show that in the absence of forcing and dissipation it shows quasi-periodic motion. In Section~\ref{forced} we analyze the system behavior in the presence of sinusoidal forcing and show that it can exhibit strange nonchaotic behavior. In Section~\ref{dissipation} we study the behavior of the system with dissipation as an open quantum system using the quantum Langevin equation and show that it can exhibit chaotic behavior. In Section~6 we conclude.

\section{The impact oscillator\label{impact}}

 In order to investigate whether quantum systems can manifest complex dynamics, we consider the quantum analog of a classical system, the impact oscillator.

\subsection{The Classical Impact Oscillator}

\begin{figure}[tbh]
    \centering
        \includegraphics[width=0.35\columnwidth]{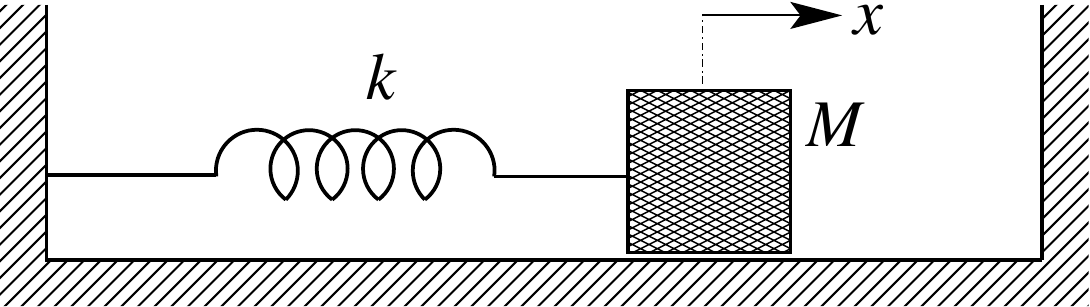}  \caption{Schematic diagram of a simple impact oscillator.}
    \label{fig:schematic}
\end{figure}

The system consists of a mass-spring arrangement in which the mass can collide with a rigid wall (schematic shown in Fig.~\ref{fig:schematic}). In the absence of any external forcing, the potential function is described by:
\begin{equation}
V(x)=\begin{cases}
\frac{1}{2} k x^2  & \text{if } x<x_w\\
\infty  & \text{if } x \ge x_w
\end{cases}
\label{potential1}
\end{equation}
where $x$ is the deviation from the relaxed position of the spring and $x_w$ is the position of the wall. 

\begin{figure}[tbh]
    \centering
    \includegraphics[width=0.3\columnwidth]{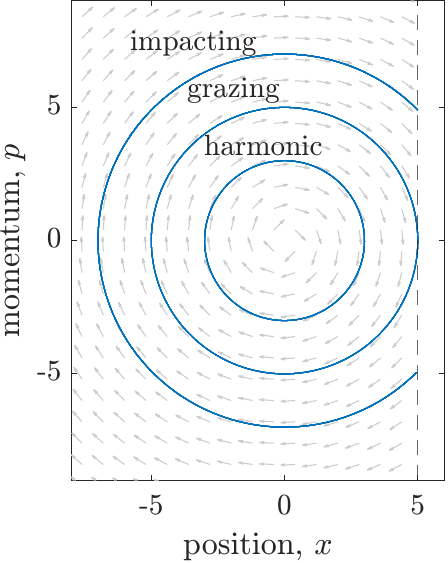}
    \caption{Phase space trajectories of the simple impact oscillator. }
    \label{fig:trajs}
\end{figure}

For this system, each trajectory in the phase space is periodic, with the {\em grazing} orbit dividing the phase space into two dynamically distinct regions (Fig.~\ref{fig:trajs}).

\subsection{Quantum Impact Oscillator}

\begin{figure}[tbh]
    \centering
    \includegraphics[width=0.4\linewidth]{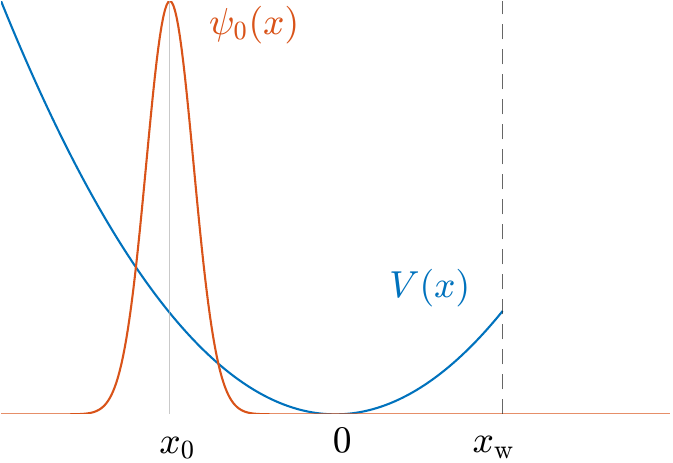}   
    \caption{The impact oscillator potential (blue) with a Gaussian wavepacket (red) as the initial condition. }
    \label{fig:init-pot}
\end{figure}
The quantum analog of this system would be a particle confined by the same potential function (\ref{potential1}) as shown in
Fig.~\ref{fig:init-pot}. The dynamics of the particle would be given by the evolution of the wavefunction $\Psi(x,t)$ following the Schr\"odinger equation
\begin{equation}
    i\hbar \frac{\partial}{\partial t} \Psi({x}, t) = \left(-\frac{\hbar^2}{2m} \frac{d^2}{dx^2} + V(x)\right) \Psi({x}, t).\label{schr}
\end{equation}

The parameters were taken as $k = 1$, $m = 1$. We have used natural units with
$\hbar = 1$.
There are several techniques to numerically solve the Schr\"odinger equation. We have used the Numerov-Cooley scheme \cite{numerov1933publs,cooley1961improved} and the results were verified by direct diagonalization of the Hamiltonian matrix \cite{izaac2018computational}.

The simulation starts from an initial condition, and in the quantum system, it would be the initial wavefunction. We choose a Gaussian wavepacket with variance $\hbar\over 2\sqrt{km}$ that has a minimum uncertainty, and shift the mean position to $x=-5$, which is equivalent to releasing the mass from a stressed position of the spring corresponding to the classical grazing condition ($x_w=5$).

‌\section{Dynamics of the quantum impact oscillator\label{unforced}}

\begin{figure}
    \centering
    \includegraphics[width=0.8\linewidth]{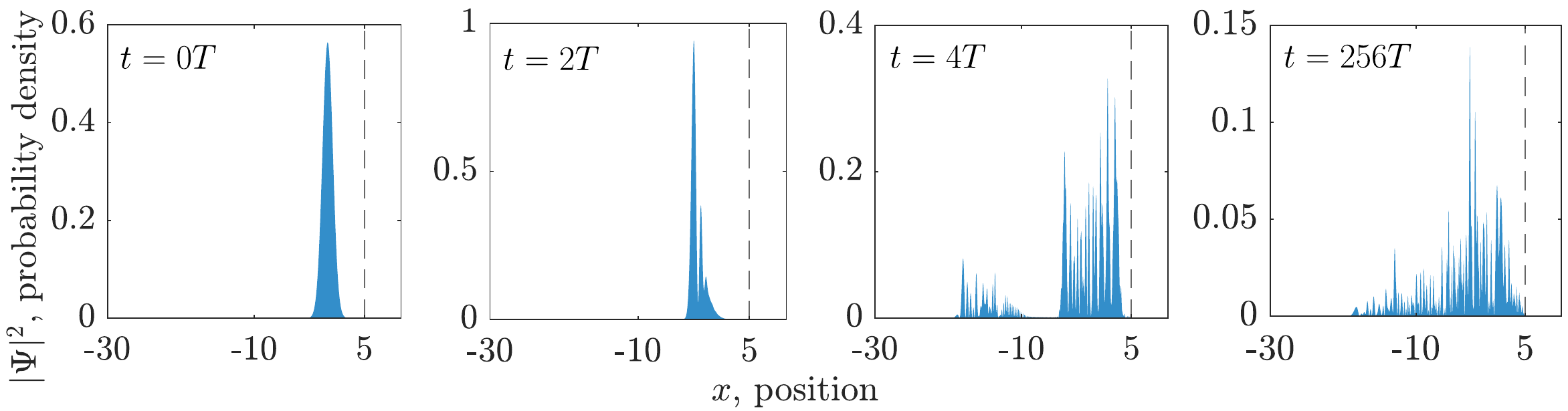}
    \caption{Snapshots of the evolution of the probability density function for the grazing initial conditions with $A_f=3.0905$  $T=2\pi/\omega_f$ is one time period of the forcing.}
    \label{fig:snaps-prob-den}
\end{figure}

\begin{figure}[tbh]
    \centering
    \includegraphics[width=0.4\columnwidth]{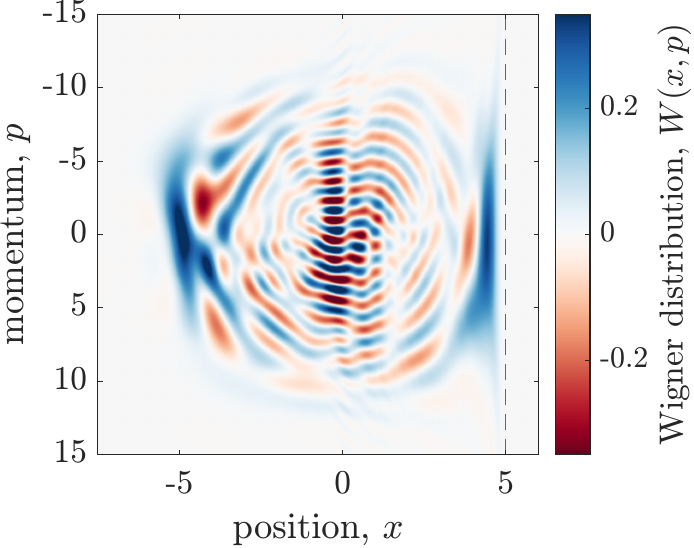}
    \caption{A snapshot of the evolving Wigner distribution for the grazing condition at $t=100$.}
    \label{fig:Wigner_grid}
\end{figure}

For the quantum impact oscillator, when the wall is placed far away, the dynamics of the wavefunction is similar to that of a harmonic oscillator (it oscillates periodically around the mean position). But it 
becomes aperiodic as the wall approaches the classical grazing condition. 
A way of representing the dynamics of a quantum system in the phase space is provided by the Wigner distribution function:
\begin{equation}
W(x, p) = \frac{1}{\pi \hbar} \int_{-\infty}^{\infty} \Psi^*\left(x + y\right) \Psi\left(x - y\right) e^{2ipy/\hbar} \, dy
\end{equation}
It gives a quasi-probability distribution over the phase space.  We found that near the grazing condition, the Wigner distribution evolves aperiodically with definite patterns. Fig.~\ref{fig:Wigner_grid} shows a snapshot of the evolving distribution. The occurrence of negative values, implying non-classical behavior, is noticeable.

\begin{figure}
    \centering
    
    \includegraphics[width=0.75\columnwidth]{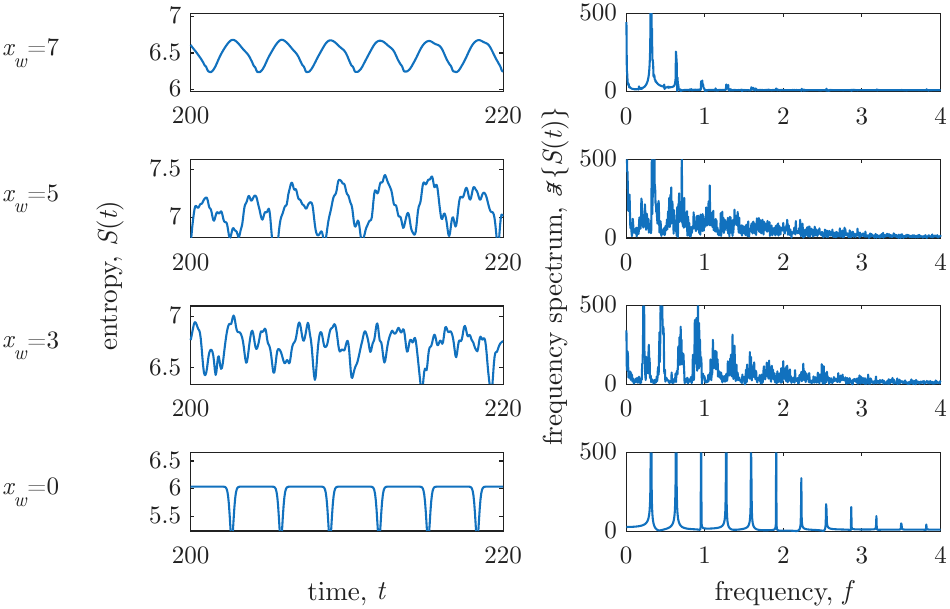} \\
    {\footnotesize \hspace{2.5cm} (a) \hspace{5.5cm} (b)}
    \caption{(a) Entropy of the probability density $|\psi(t)|^2$ versus time for different wall positions. (b) Frequency spectrum of the entropy time-series. }
    \label{fig:entropy}
\end{figure}

In order to find out the qualitative character of the dynamics, we need to convert the dynamics of the complex-valued wavefunction into a real-valued time series.
The entropy of the probability density \citep{shannon1948mathematical} 
\begin{equation}
    S(t) = -\int |\Psi(x,t)|^2 \log(|\Psi(x,t)|^2) dx
\end{equation}
 generates such a time series (see Fig.~\ref{fig:entropy}a). It shows that when the wall is far away the behavior is periodic. As the wall is moved closer ($x_w=6$), the waveform becomes aperiodic. In the classical grazing condition ($x_w=5$), the aperiodic nature is quite pronounced. When the wall is moved to the middle of the potential function, i.e., when the system is a `half-harmonic oscillator' ($x_w=0$), the waveform is again periodic. 

Fig.~\ref{fig:entropy}b shows the frequency spectrum of this time series. It shows an increasing number of discrete peaks as the wall approaches the grazing condition, indicating a complex quasiperiodic motion. 
This behavior contrasts sharply with that of the classical system, which remains periodic for all wall positions.

When the time series corresponding to different wall positions were subjected to the 0-1 test \cite{gottwald2004new}, they returned values close to zero, indicating that the visible aperiodicity in the time series is due to quasiperiodicity rather than chaos.
This result is consistent with the linear nature of quantum evolution. 

\section{The impact oscillator with forcing
\label{forced}}

\begin{figure*}[t]
    \centering
    \includegraphics[width=\linewidth]{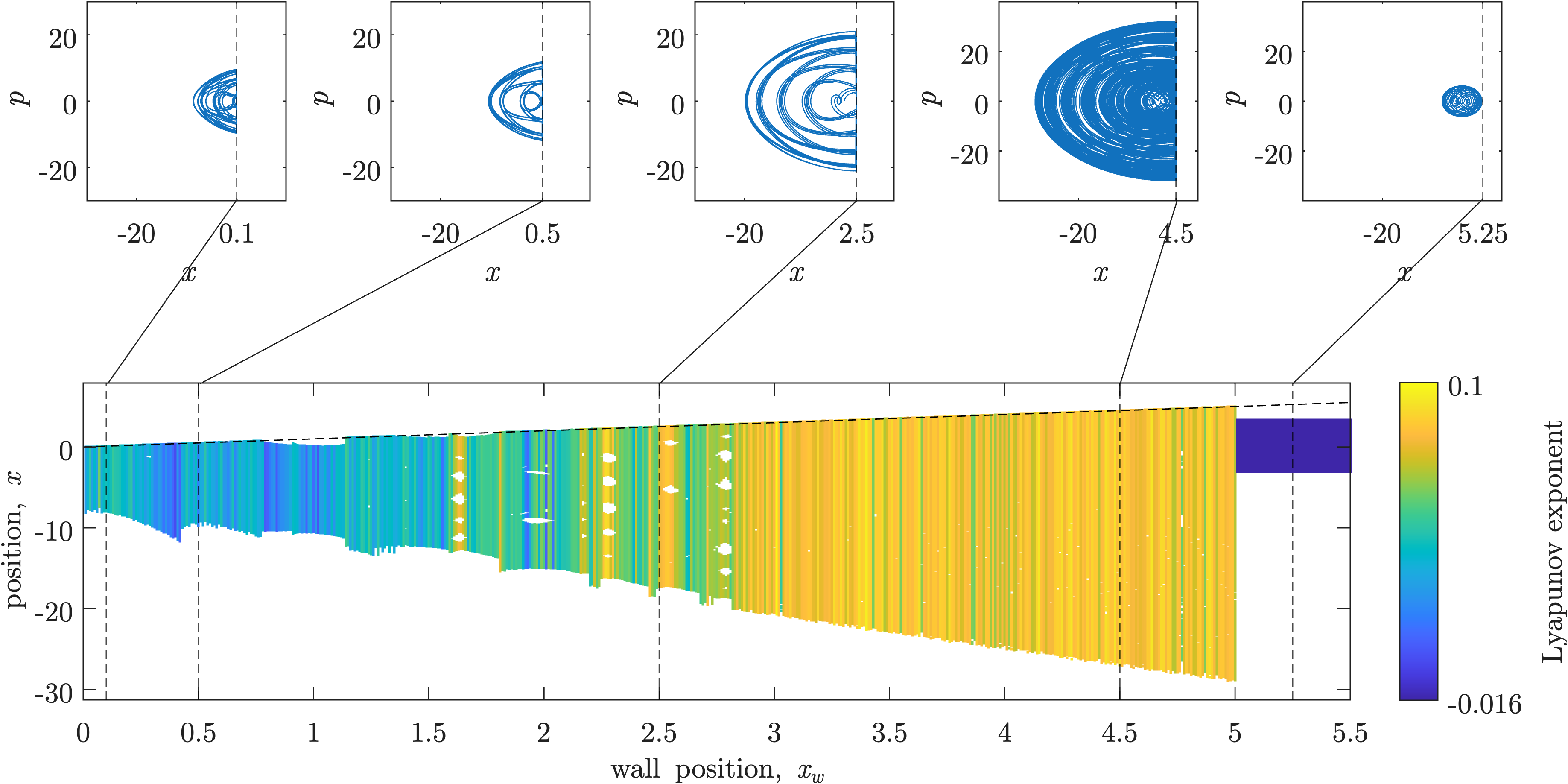}
    \caption{Bifurcation diagram of the classical forced impact oscillator as a function of wall position. Color indicates the maximal Lyapunov exponent.}
    \label{fig:bifclass}
\end{figure*}

When sinusoidal forcing $f(t) = A_f \sin(\omega_f t)$ is applied to the mass, the classical impact oscillator shows (Fig.~\ref{fig:bifclass}) a sudden transition to a chaotic oscillation of large amplitude when the mass grazes the wall \citep{nordmark1991non, ing2008experimental, bernardo2008piecewise}. The system exhibits windows of chaos, quasiperiodicity, and periodicity depending on the wall position and forcing parameters. 

\subsection{The quantum analog with forcing}

In this case, the evolution of the wavefunction is given by the same Schr\"odinger equation (\ref{schr}), with the potential function changed to 
\begin{equation}
V(x)=\begin{cases}
\frac{1}{2} k x^2 + x \: A_f \sin(\omega_f t)  & \text{if } x<x_w\\
\infty  & \text{if } x \ge x_w
\end{cases}
\label{potential2}
\end{equation}

\noindent Solving the Schr\"odinger equation for a time-dependent Hamiltonian is a critical task in quantum mechanics. We adopt the Commutator-free Exponential Time-Propagators (CFETs) method
\cite{ALVERMANN20115930} to solve this system, which is outlined in Appendix~A.


\subsection{Dynamics of the quantum analog}

What are the behaviors exhibited by the forced impact oscillator in the quantum regime?

The initial state is assumed to be a Gaussian wavepacket with mean at $x = 0$ and variance ${\hbar}/(2\sqrt{km})$. The natural frequency of the harmonic oscillator without the wall is 1, and the forcing frequency is considered an irrational number $(\sqrt{5}+1)/{2}$.

\begin{figure}[tbh]
    \centering
   
    \includegraphics[width=0.7\columnwidth]{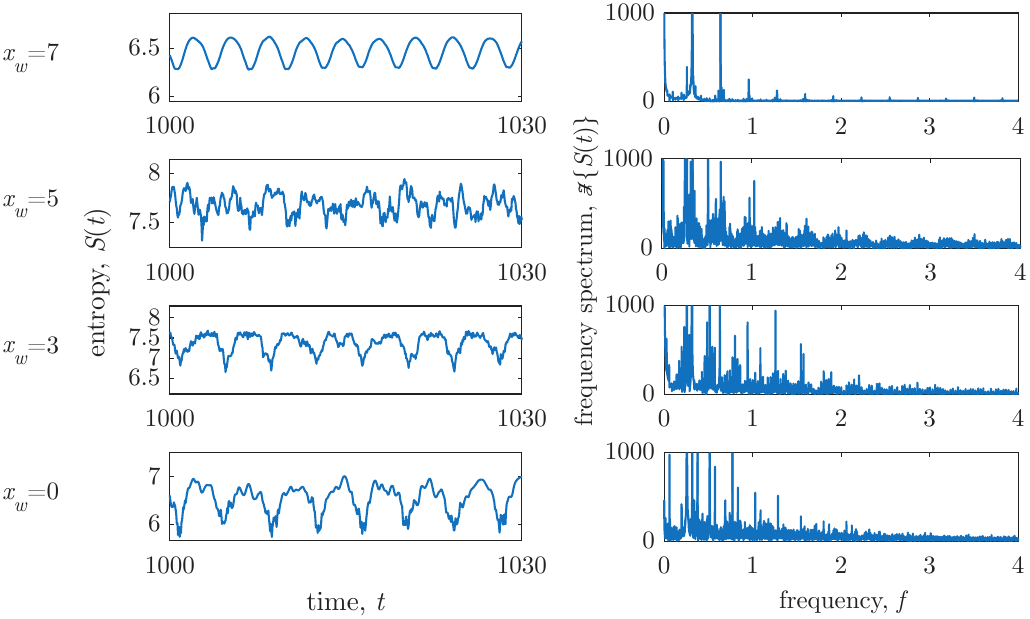}\\
     {\footnotesize \hspace{2cm} (a) \hspace{2in} (b)}
    \caption{(a) Time-series of the entropy of the probability distribution. (b) Fourier spectra of the entropy time-series for different positions of the wall. }
    \label{fig:entFTforced}
\end{figure}

Upon solving the equation numerically, we found that the forced quantum impact oscillator reveals more exotic behavior (Fig.~\ref{fig:entFTforced}a). 
As the wavefunction is an extended entity in space, the grazing condition cannot be defined as in the classical case. To find the grazing
condition, we remove the wall and observe the dynamics of the coherent state
in the (now) harmonic potential subjected to the same sinusoidal driving. We
allow the time evolution to reach a steady state of oscillation. The condition
for which the peak of the probability distribution $|\Psi|^2$ just reaches the position where the
wall would have been, is taken to be the grazing condition for the quantum impact oscillator. For $x_w = 5$ and $\omega_f=(\sqrt{5}+1)/{2}$, we find the condition for grazing as $A_f = x_w  m \omega_0 \left|\omega_f - \omega_0 \right| \approx 3.09$ where $\omega_0=\sqrt{k/m}$ is the natural frequency. In subsequent simulations, this value of $A_f$ has been set, so
$x_w = 5$ can be taken as the grazing condition.

We notice the following aspects. These were first reported in \cite{acharya2023sna}; here we report the results obtained using the more accurate CFETs algorithm.
\begin{figure}
     \centering
     \includegraphics[width=0.5\columnwidth]{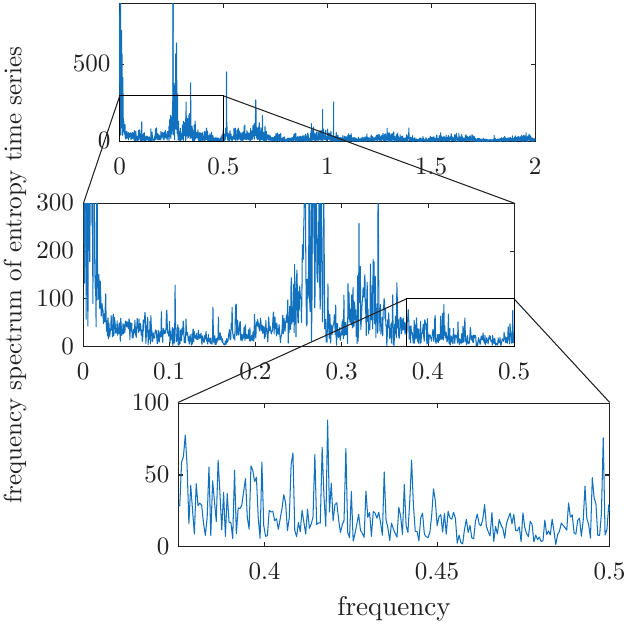}
     \caption{Two consecutive zooms of the frequency plot corresponding to Fig.~\ref{fig:entFTforced}, the case of $x_w = 5$. }
     \label{fig:FT_zoom}
 \end{figure}
\begin{description}\setlength{\itemsep}{1em}   
\item[Dense discrete spectrum:] The Fourier spectrum of the entropy time series (Fig.~\ref{fig:entFTforced}b) shows that, as the wall position approaches the grazing condition,  a large number of discrete peaks appear.  Repeated zooms of the spectrum shown in Fig.~\ref{fig:FT_zoom} reveal a dense set of discrete frequency peaks. This could be indication of a spectrum with fractal character (singular continuous spectrum). It has been shown \citep{bezhaeva1996example} that the appearance of a dense discrete spectrum is a characteristic of strange nonchaotic orbits.

\item[Power-law character in the spectral distribution function:] To further investigate the issue, we obtain the spectral distribution function $N(\sigma)$: the number of frequency components above a threshold $\sigma$. Earlier research \citep{pikovsky1995characterizing,prasad2001strange} has shown that, if the time series comes from a strange non-chaotic orbit, then the $N(\sigma)$ versus $\sigma$ plot would show a power-law scaling. To verify, we plot $N(\sigma)$ versus $\sigma$ in log-log scale (Fig.~\ref{fig:spectlyap}a). We find that it exhibits power-law scaling with an exponent of $-2$.

\begin{figure}
    \centering    \includegraphics[width=0.28\columnwidth]{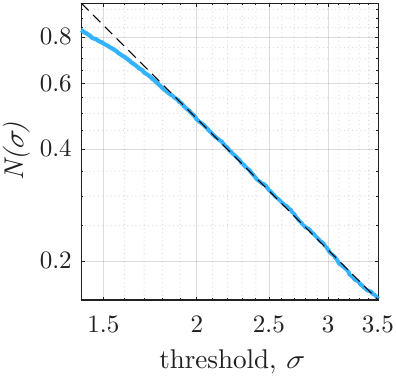}\hspace{0.5cm}
\includegraphics[width=0.3\columnwidth]{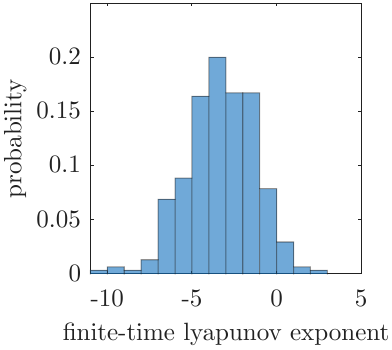}
    \hspace{0.3cm}
    \includegraphics[width=0.285\columnwidth]{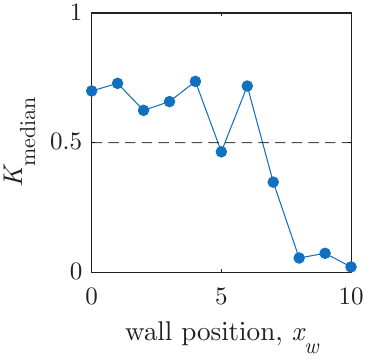}
    \\
     {\footnotesize \hspace{0.5cm} (a) \hspace{5cm} (b) \hspace{5cm} (c)}
    \caption{(a) Spectral distribution function, i.e., the number of peaks above threshold $\sigma$ versus the threshold for the frequency spectrum of the entropy time-series for the forced quantum impact oscillator in log-log scale.
    (b) Distribution of finite-time Lyapunov exponents of the entropy time-series. Presence of positive parts indicates fractal structure of the underlying dynamics. (c) The variation of $K_{\rm median}$, the metric from the modified 0-1 test, as the wall position is varied.}
    \label{fig:spectlyap}
\end{figure}

\item[Local instabilities:]
It is known that, even though the largest Lyapunov exponent for a strange non-chaotic orbit is negative, the dynamics can be locally unstable. This
is captured by the finite-time Lyapunov exponents \citep{grassberger1988scaling, kapitaniak1995distribution}. For the quantum forced impact oscillator, the distribution of finite-time Lyapunov exponents (Fig.~\ref{fig:spectlyap} (b)) includes positive values, indicating local instability despite the globally non-chaotic nature \citep{prasad1999characteristic}.
\begin{figure}[h!]
    \centering
    \begin{subfigure}[b]{0.3\textwidth}
        \centering
        \includegraphics[width=\textwidth]{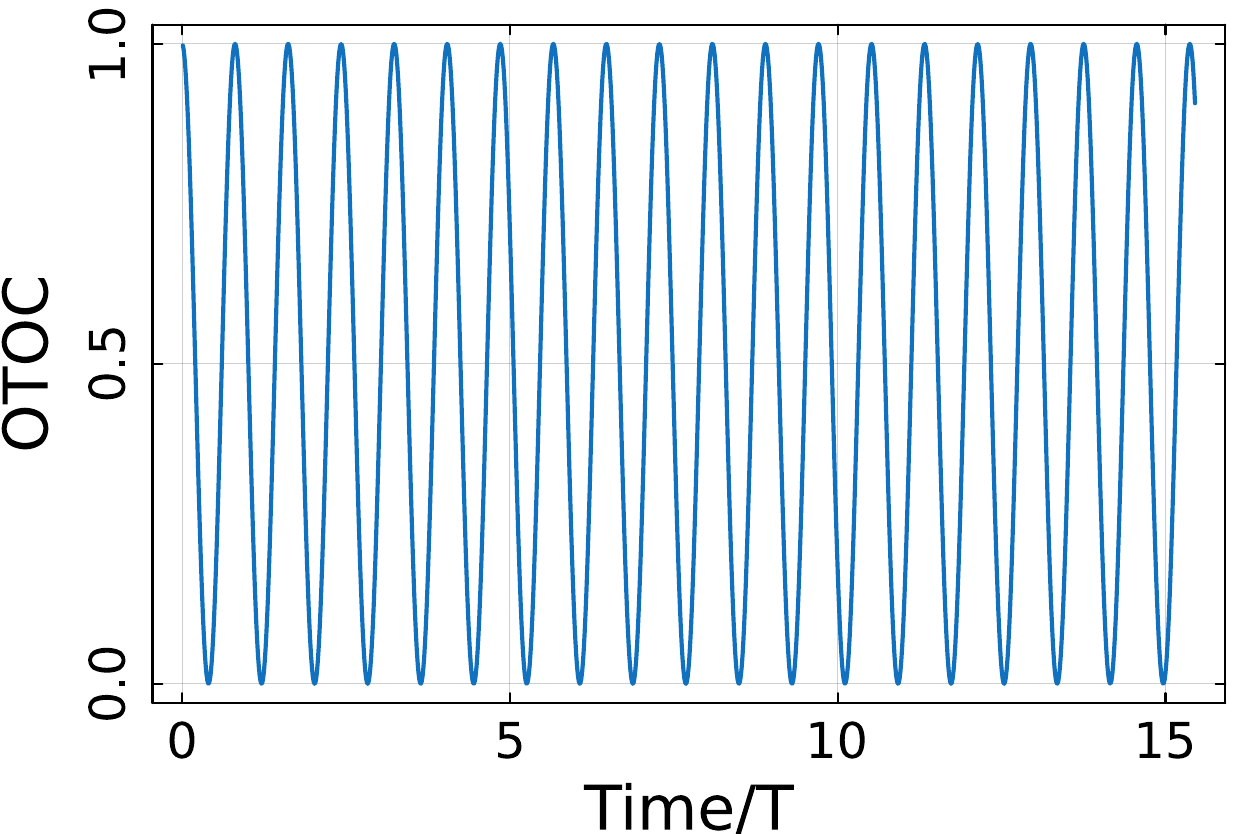}
        \caption{}
        \label{fig:otoc_inf}
    \end{subfigure}
    \hfill
    \begin{subfigure}[b]{0.3\textwidth}
        \centering
        \includegraphics[width=\textwidth]{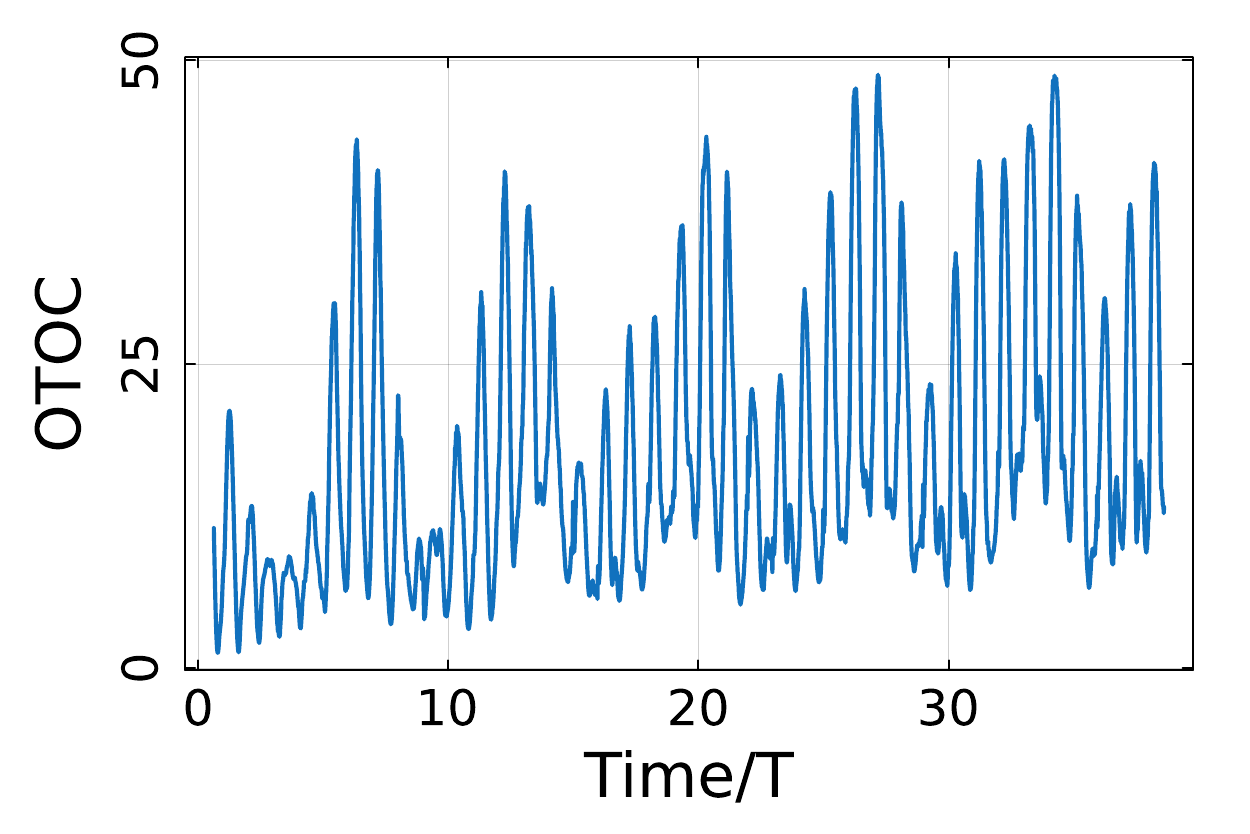}
        \caption{}
        \label{fig:otoc0}
    \end{subfigure}
    \hfill
    \begin{subfigure}[b]{0.3\textwidth}
        \centering
        \includegraphics[width=\textwidth]{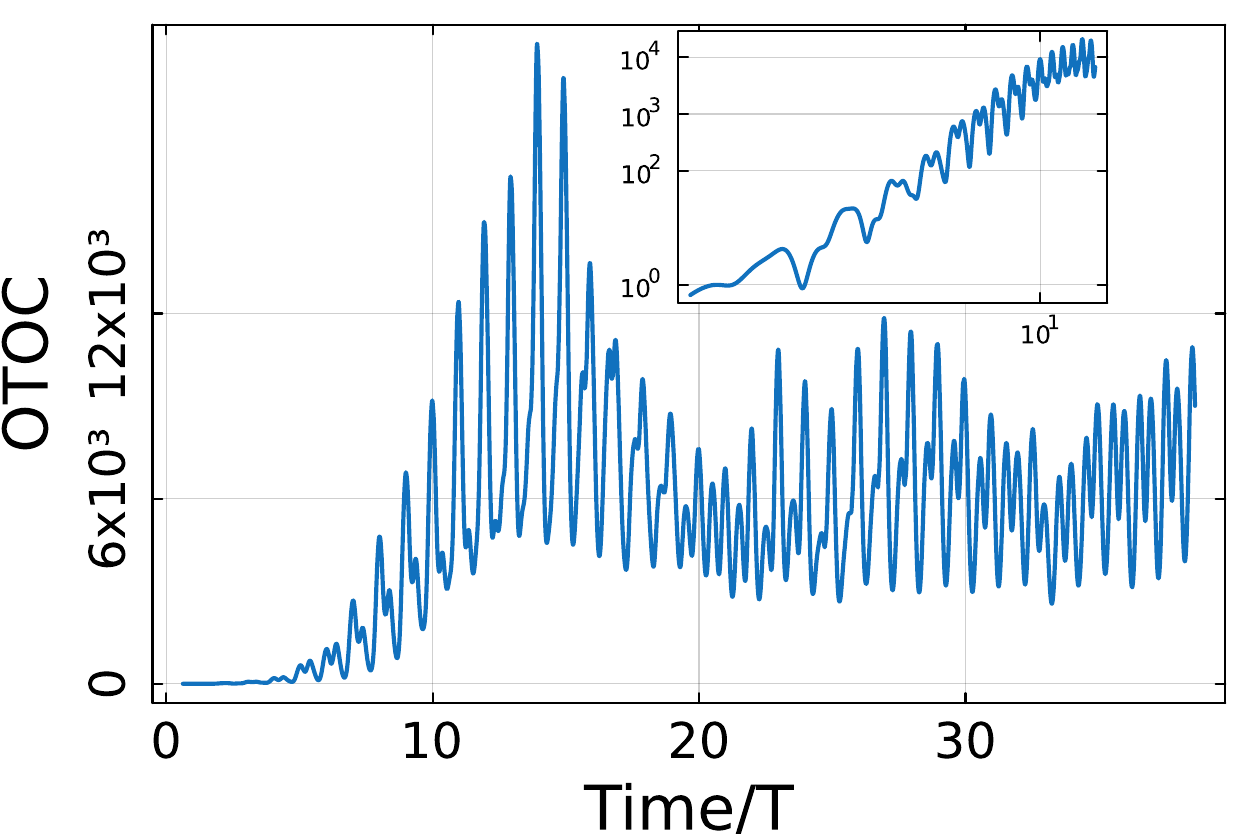}
        \caption{}
        \label{fig:otoc5}
    \end{subfigure}
    \caption{OTOC temporal behavior for $\beta=0.5$ at three different wall positions: (a) $x_{w}=\infty$ (harmonic case), (b) $x_{w}=1.2$, and (c) $x_{w}=5.0$. Panel (a) shows periodic behavior, (b) displays a slow increasing trend, and (c) exhibits initial power-law growth followed by saturation. The inset in (c) confirms the power-law growth via a log-log plot. $T$ refers time period of forcing.}
\end{figure}

\item [OTOC:] 
We compute the out-of-time-order correlator (OTOC) using the method described in \cite{hashimoto2017out}. The OTOC, defined by (\ref{otoc1}), is a quantity that helps track how quickly small perturbations spread through a quantum system over time. We take the two operators as $x(t)$ and $p(0)$ in Heisenberg's picture (the position and momentum operators at times $t$ and 0) so that the OTOC becomes $C_T(t) = -\left\langle [x(t), p(0)]^2 \right\rangle,$
where the brackets $\langle \cdot \rangle$ represent a thermal or quantum average.

To make this more concrete, we express the thermal OTOC using the energy eigenstates of the system:
\begin{equation}
C_T(t) = \frac{1}{Z} \sum_n e^{-\beta E_n} c_n(t), \quad \text{with } c_n(t) \equiv -\langle n | [x(t), p]^2 | n \rangle,
\end{equation}
where $Z$ is the partition function, $\beta=\frac{1}{k_B T}$ represents the inverse temperature as we consider $k_B=1$, and $E_n$ and $\ket{n}$ are the energy eigenvalues and eigenstates of the system's time-independent Hamiltonian.

\begin{figure}
    \centering
        \includegraphics[width=0.6\columnwidth]{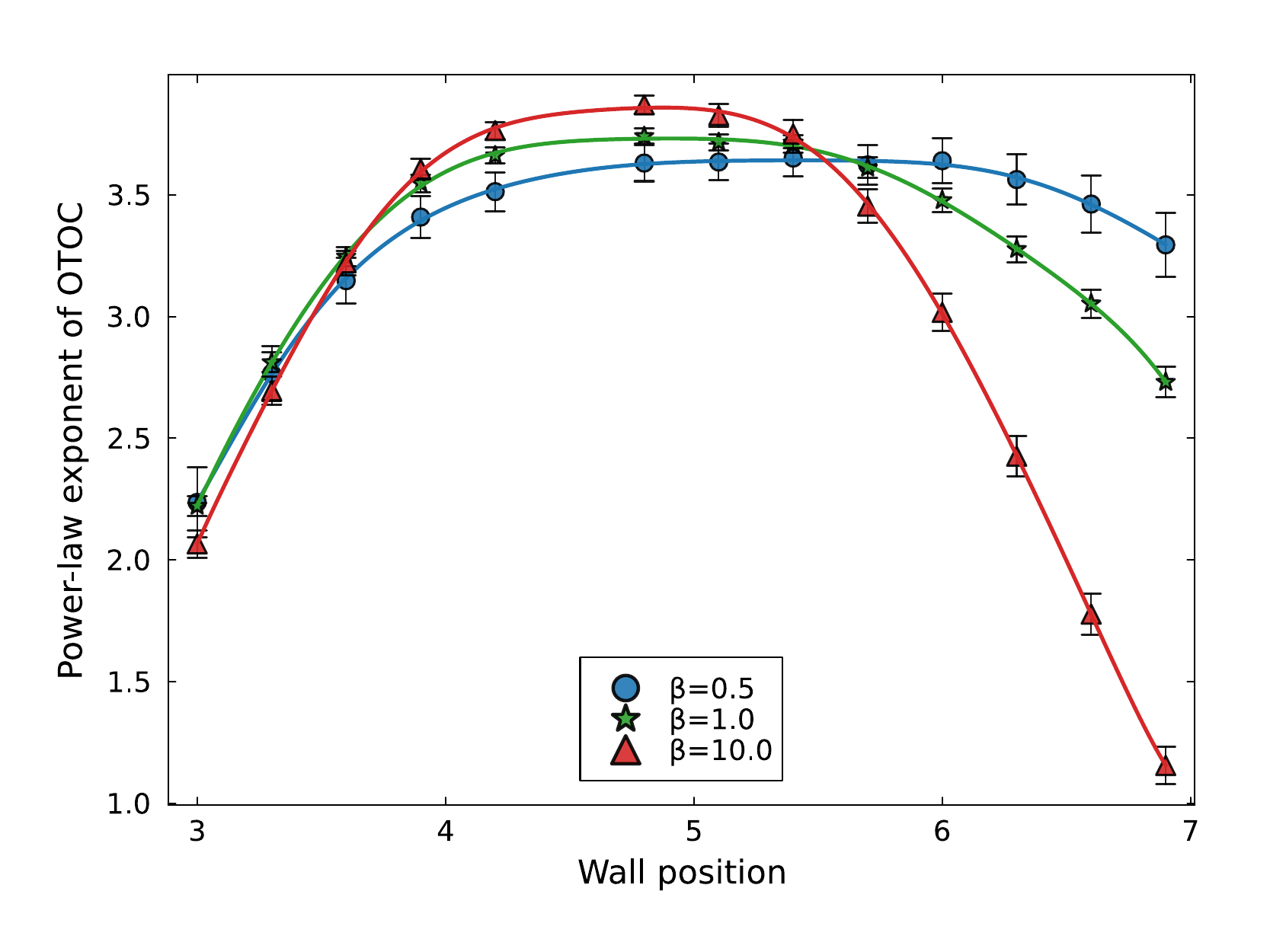}
    \caption{Power-law exponent of initial OTOC growth with respect to wall position $x_{w}$ for three different temperatures $\beta=0.5, 1.0, 10.0$. A polynomial fit has been used to estimate the trend. $\beta=0.5$ is shown in blue with circle marker, $\beta=1.0$ in green with star marker, and $\beta=10.0$ in orange with triangle marker. The error bars correspond to the standard error when fitting the initial OTOC growth with a power-law.}
    \label{fig:power_otoc}
\end{figure}

We evaluate the OTOC for different positions of the wall, $x_w$, ranging from $0$ to $10$ and $\infty$, using 600 eigenstates at various inverse temperatures ($\beta=0.5$, $1.0$, and $10.0$), with forcing amplitude $A_f=3.0905$ and frequency $\omega_f=(\sqrt{5}+1)/{2}$. For the harmonic case ($x_w = \infty$), the OTOC remains periodic with significantly low amplitude (see Fig.~\ref{fig:otoc_inf}). In contrast, for $x_w = 0$ to $3.0$, it shows a slight increasing trend (Fig.~\ref{fig:otoc0}). 

For near-grazing impacts ($3.0 < x_w < 7.0$), the OTOC exhibits an initial power-law growth followed by saturation and bounded oscillations (Fig.~\ref{fig:otoc5}), similar to behavior reported in quasiperiodically driven systems such as the Aubry–André model~\cite{riddell2020out} and also other non-chaotic systems such as the Ising Floquet system \cite{shukla2022out}. We fit the early-time OTOC growth to a power-law across the wall positions for the chosen temperatures. The exponent, quantifying the rate of information scrambling, increases with $x_w$ approaching grazing conditions and then decreases. This peak shifts to smaller $x_w$ as temperature is lowered (i.e., higher $\beta$) (see Fig.~\ref{fig:power_otoc}). Notably, we do not observe any exponential growth, confirming the absence of quantum chaos despite the presence of grazing impact.

\item[0-1 test:]
It is known that, in a 0-1 test, a strange non-chaotic time series returns intermediate values between 0 and 1 \citep{kim2014numerical, gopal2013applicability}.  A few modifications of the 0-1 test make it more reliable for identification of strange non-chaotic dynamics, viz. addition of a small noise term \cite{dawes20080}, and fixing the value of $c$ (a randomly chosen value in the original 0-1 test to prevent resonances) to suitable irrational numbers \cite{gopal2013applicability}.
For the quantum system,  the modified 0-1 test applied to the entropy time series at the grazing condition yielded $K_{\rm median}=0.46$,  indicating strange non-chaotic dynamics (Fig.~\ref{fig:spectlyap}(c)).

\item[Bifurcation diagram:]
We plot the quantum bifurcation diagram by taking stroboscopic snapshots of the quantum probability density function and averaging it over a considerable number of periods. What we obtain is a distribution where the particle is more likely to be found. Then we plot this distribution, using color to represent high and low values,  against a parameter, the wall position in our case. The quantum bifurcation diagram (Fig.~\ref{fig:Bifurcation_FFT}) shows a dramatic increase in the complexity of probability density patterns around the grazing condition, reminiscent of the large-amplitude chaotic orbits in the classical system. 
\begin{figure}
    \centering
        \includegraphics[width=0.5\columnwidth]{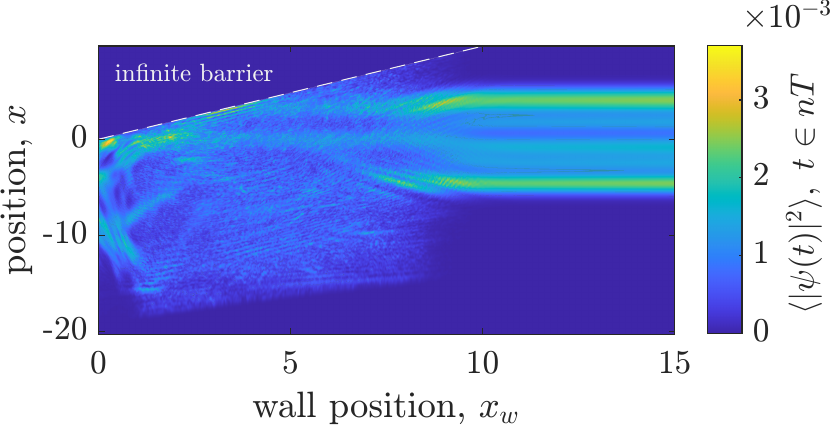}
    \caption{The quantum bifurcation diagram for the forced impacting system.}
    \label{fig:Bifurcation_FFT}
\end{figure}
\end{description}

Based on the above observations, we conclude that the quantum forced harmonic oscillator exhibits strange nonchaotic dynamics close to the grazing condition.

\section{The quantum forced impact oscillator with dissipation\label{dissipation}}

To model the system with dissipation, we use the quantum Langevin equation. 
Specifically, we use the semiclassical formalism developed in \cite{barik2003numerical, banerjee2004solution}, which enables us to work with the quantum expectation values of position and momentum, along with their fluctuations. A brief description of the method is given in Appendix B.

\begin{figure}[hbt!]
    \centering
    \includegraphics[width=0.4\linewidth]{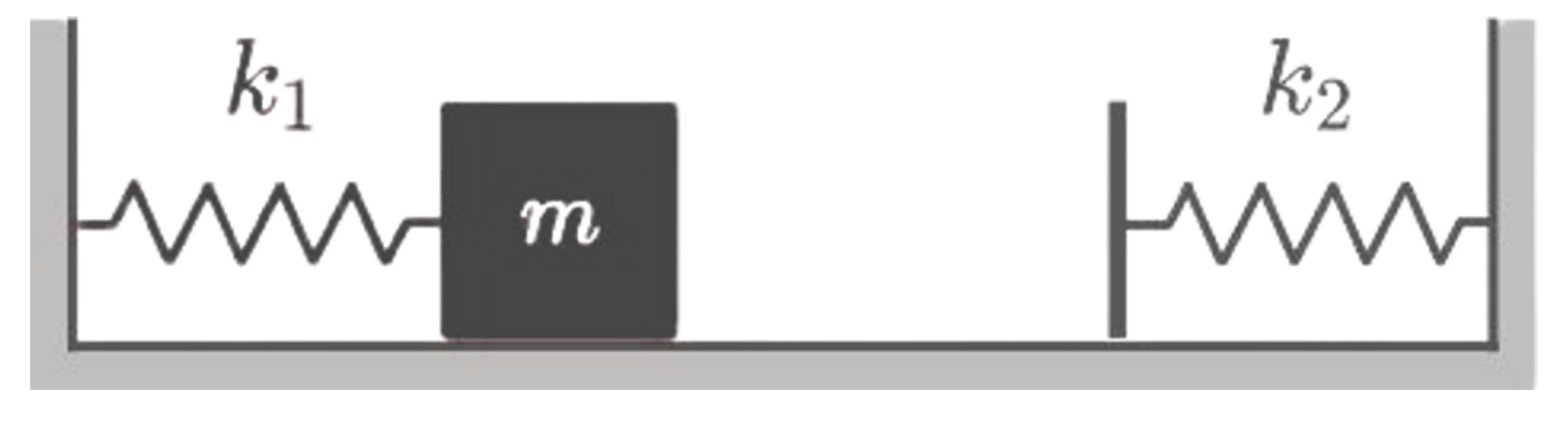}
    \caption{Schematic diagram of a soft impact oscillator}
    \label{fig:soft_schem}
\end{figure}

In the hard-impact oscillator, the potential function (\ref{potential1}) has a derivative discontinuity, which causes all subsequent derivatives to be discontinuous. Since such terms appear in the model (see eqn. \ref{eqn:Q}), this model cannot be used.

To address this, we consider a soft impact oscillator, where the rigid wall is replaced by a stiff, yet smooth, spring-like potential as shown schematically in Fig.~\ref{fig:soft_schem}. The mathematical form of the potential is given by (\ref{eqn: pot}):
\begin{equation}
V(x,t)  = \left\{\begin{array}{ll}
\frac{1}{2} k_1 x^2 + x A_f \cos(\omega_f t), & \;\;\text{if } \;\;x < x_w  \\ \\
      \frac{1}{2} k_1 x^2 + \frac{1}{2} k_2 (x - x_w)^2 + x A_f \cos(\omega_f t), & \;\;\text{if }\;\; x \ge x_w,
     \end{array}
     \right.
\label{eqn: pot}
\end{equation}

\medskip
\noindent where we take $k_1=k$ and $k_2=Ak$, and $k$ is the spring constant of the main oscillator, $A$ is a stiffness factor for the additional spring that activates beyond the position of the wall $x_w$, and $A_f$ and $\omega_f$ are the amplitude and frequency of external forcing, respectively.

In the classical domain, the soft impact oscillator exhibits rich dynamical behavior, including chaotic attractors, multistability, and grazing-induced bifurcations \cite{banerjee2009invisible,kundu2012singularities}.

To model the quantum analog of this system, the discontinuity is smoothed into a sigmoid function.
The second derivative of the potential is then modeled as follows.
\begin{equation}
V''(x) = \frac{kA}{1 + \exp\left(-c(x - x_w)\right)} + k,
\end{equation}
where the parameters $A$ and $c$ determine the stiffness and smoothness of the soft wall located at $x_w$. By integrating the sigmoid function, we have obtained the first-order derivative and its potential by choosing an appropriate integrating constant. 

\begin{figure}[tbh]
    \centering
    \begin{subfigure}[b]{0.3\linewidth}
        \includegraphics[width=\linewidth]{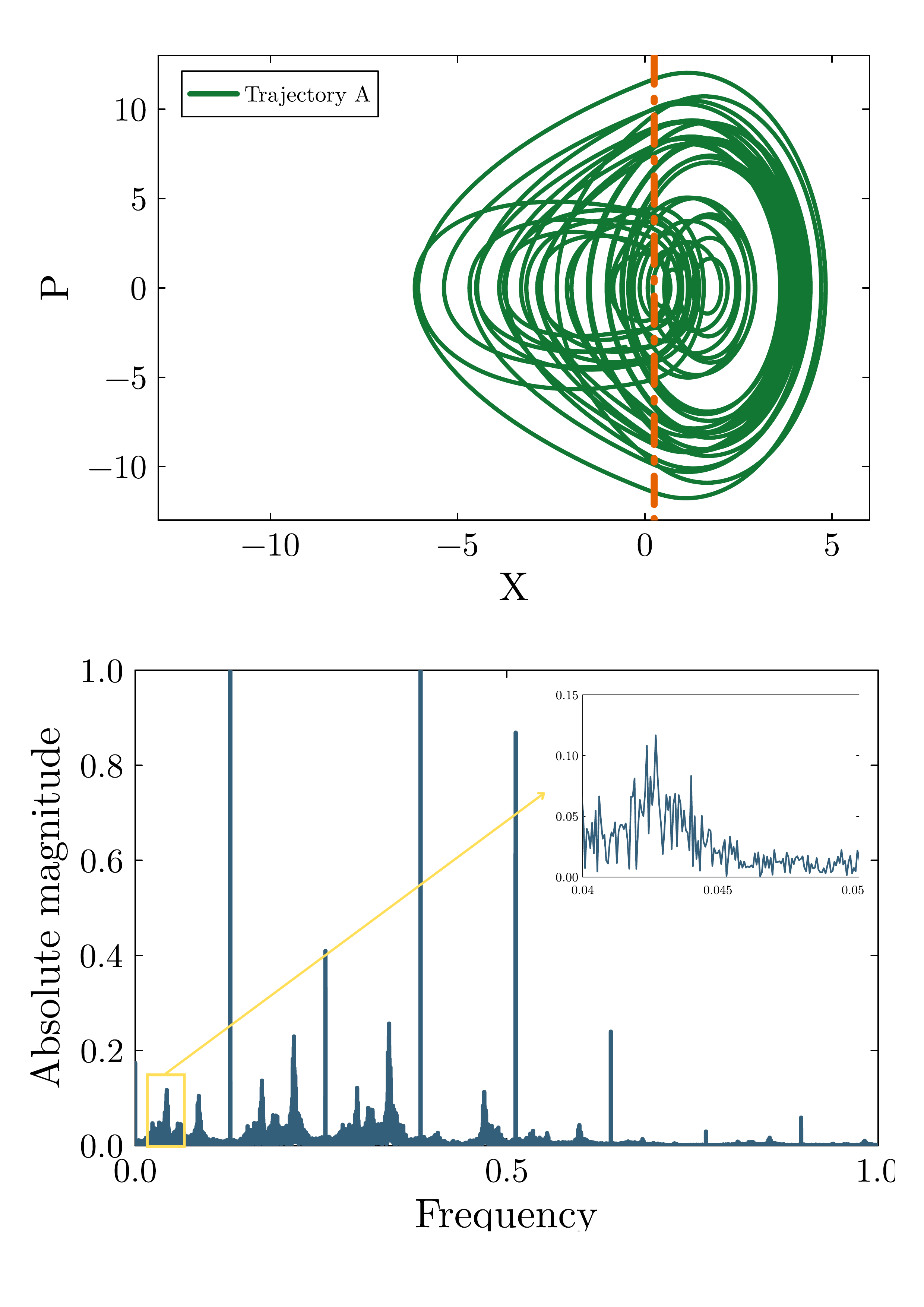}
        \subcaption{}
        \label{fig:025}
    \end{subfigure}
    \hfill
    \begin{subfigure}[b]{0.3\linewidth}
        \includegraphics[width=\linewidth]{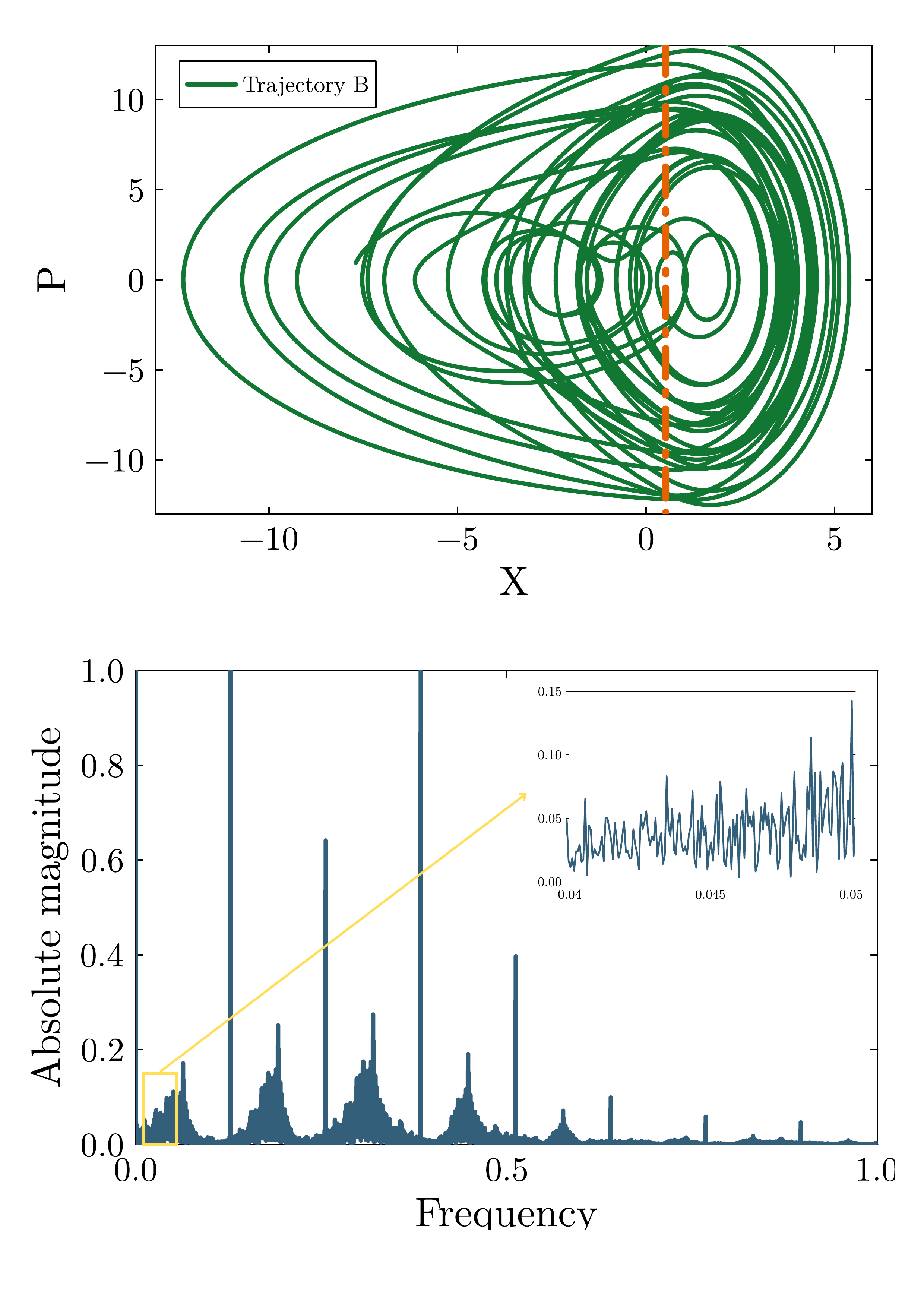}
        \subcaption{}
        \label{fig:052}
    \end{subfigure}
    \hfill
    \begin{subfigure}[b]{0.3\linewidth}
        \includegraphics[width=\linewidth]{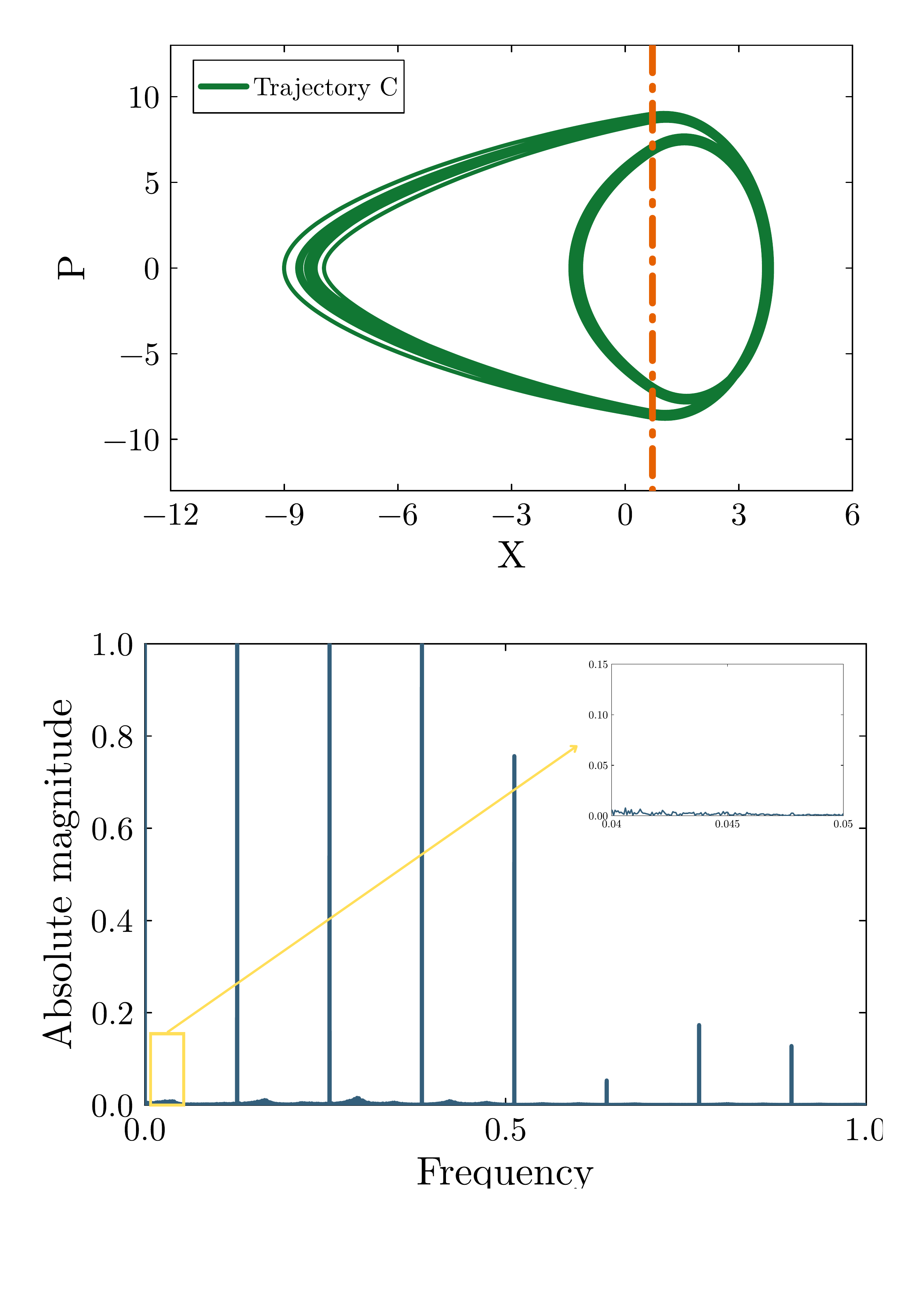}
        \subcaption{}
        \label{fig:075}
    \end{subfigure}
    \caption{Trajectories (upper panels) with wall position marked by red dash-dot line and corresponding frequency spectra (lower panels) for different wall positions at fixed $\omega_f = 0.8046$: (a) $x_w = 0.25$ (Trajectory A), (b) $x_w = 0.52$ (Trajectory B), and (c) $x_w = 0.75$ (Trajectory C). The insets highlight the spectral content. In (a) and (b), the FFTs exhibit both dominant peaks and a surrounding broadband structure, indicating chaotic motion. In contrast, (c) shows sharp, well-separated spectral lines with no surrounding broad features, characteristic of periodic dynamics. 
    \label{fig:traj_all_walls}}
\end{figure}

\begin{figure}[hbt!]
    \centering
    \includegraphics[width=0.5\linewidth]{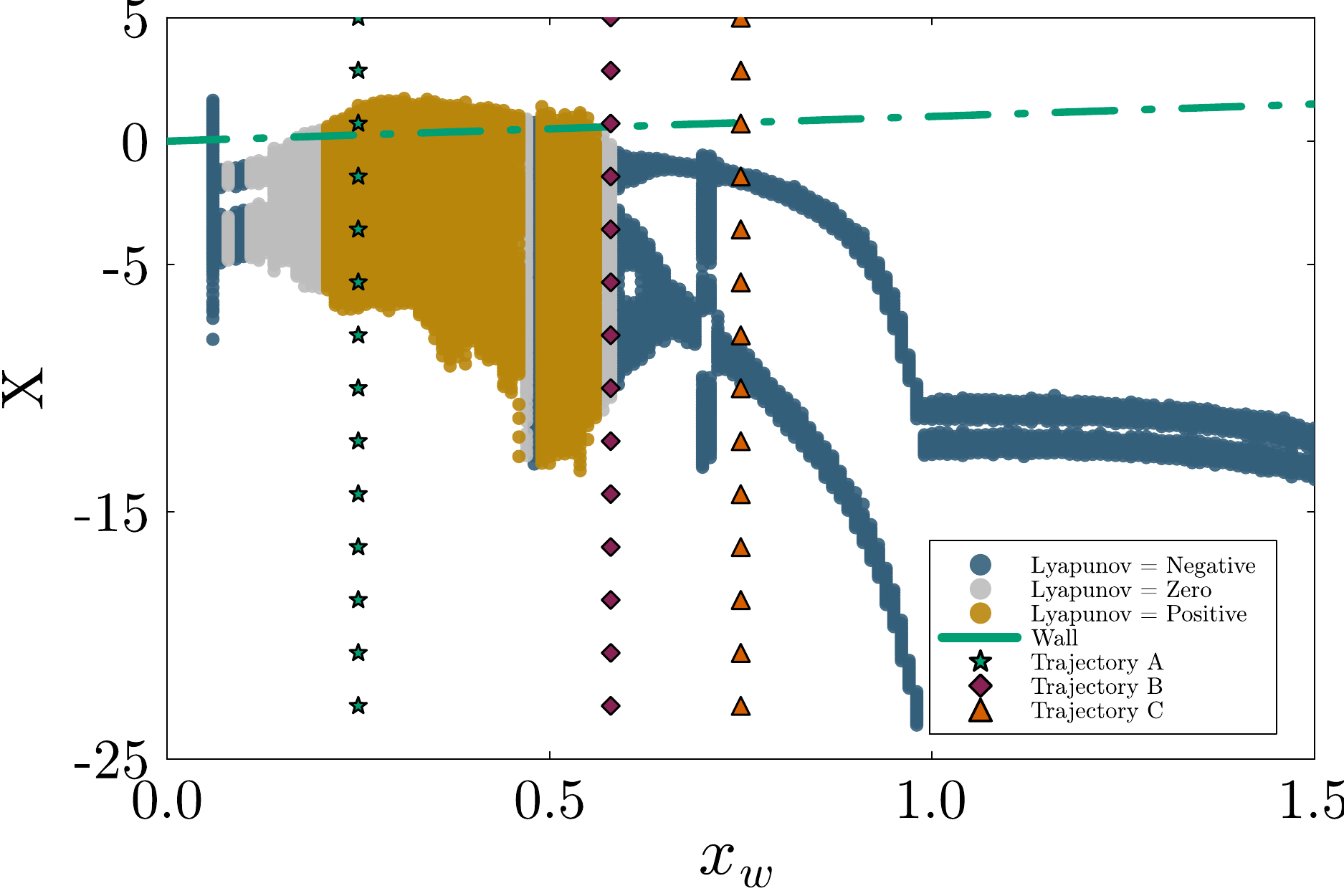}
    \caption{Quantum bifurcation diagram at $V = 0$ for varying $x_w$. Colors represent the maximal Lyapunov exponent (LE): blue (negative), grey (near-zero), ochre (positive). Vertical lines indicate wall positions: green star (Trajectory A, $x_w = 0.25$), purple diamond (Trajectory B, $x_w = 0.52$), and orange triangle (Trajectory C, $x_w = 0.75$).}
    \label{fig:bif}
\end{figure}

\begin{figure}[hbt!]
    \centering
    \includegraphics[width=0.5\linewidth]{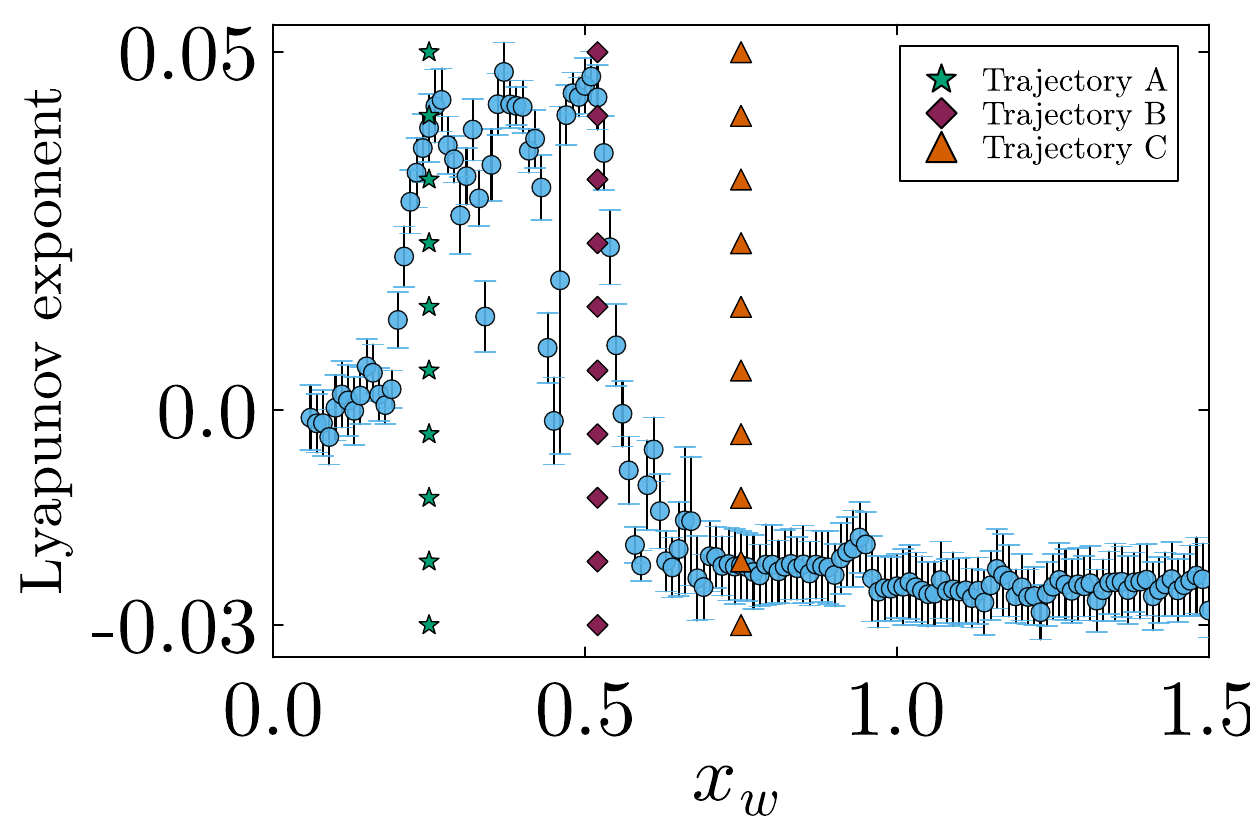}
    \caption{Mean Lyapunov exponent (sky-blue) with standard deviation error bars from 1,000 noise realisations for each $x_w$. Vertical lines mark Trajectories A ($x_w = 0.25$, green star), B ($x_w = 0.52$, purple diamond), and C ($x_w = 0.75$, orange triangle). }
    \label{fig:lyap}
\end{figure}

\begin{figure}[hbt!]
    \centering
    \includegraphics[width=0.5\linewidth]{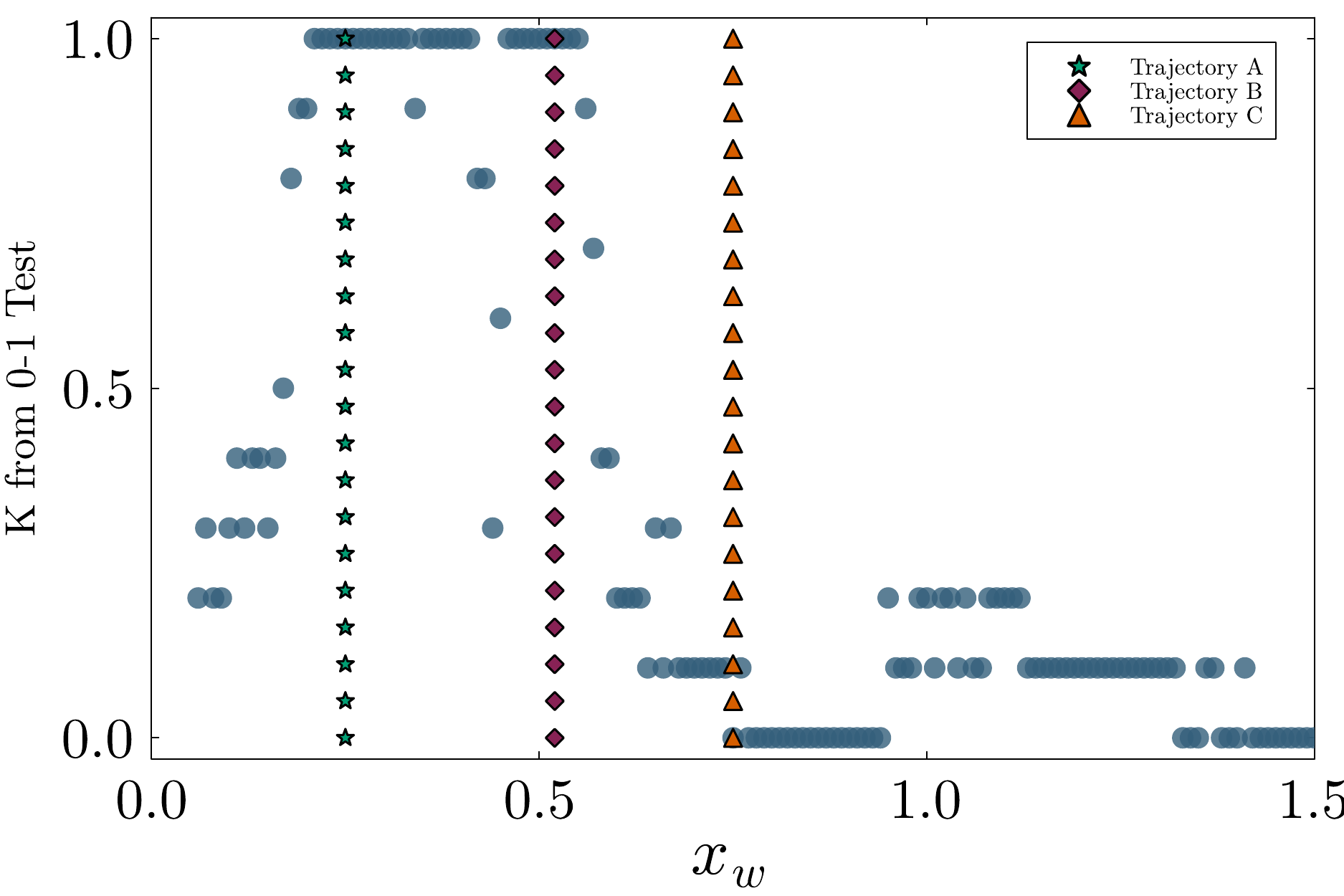}
    \caption{Results of 0–1 test as $x_w$ is varied. Values close to 1 suggest dynamical instability. Vertical lines correspond to Trajectories A (green star), B (purple diamond), and C (orange triangle). }
    \label{fig:01}
\end{figure}

\subsection{Signature of Dynamical Chaos}

We use the complex number quantum Langevin equation to simulate the dynamics, which are outlined in Appendix~B. The system parameters used in the simulation were $k = 1.0$, $A = 10.0$, $m = 1.0$, $A_f=10.0$, $\omega_f = 0.8046$. The other important parameters related to the environment are temperature ($kT$), noise correlation strength ($\Gamma$), correlation time ($\tau_c$), and the effective Planck constant ($\hbar$). We choose $kT = 0.01$, $\Gamma = 1.0$, $\tau_c = 3.0$, and $\hbar = 0.01$ for our simulation.
Fig.~\ref{fig:traj_all_walls} shows the trajectories that represent the averaged dynamics for different positions of the wall. The lower panels show the frequency spectra. The broadband structure of the trajectories $A$ and $B$ indicates chaotic behavior. In contrast, for trajectory C presented in Fig.~\ref{fig:075}, the FFT displays only sharp isolated peaks, consistent with regular periodic motion. 

By varying the wall position $x_w$, we constructed a quantum bifurcation diagram (Fig.~\ref{fig:bif}) by sampling the dynamics in the plane where the average velocity $V = 0$ . The waveforms shown in Fig.~\ref{fig:traj_all_walls} (a), (b) and (c) correspond to the parameter values marked with star, diamond, and triangle, respectively.  We observed positive Lyapunov exponents within the intervals $x_w \in [0.20, 0.43]$ and $x_w \in [0.47, 0.55]$.

Since the system is stochastic, the presence of a positive Lyapunov exponent (LE) does not, by itself, confirm chaotic behavior. Following the methodology of \cite{ruidas2024semiclassical, bagnoli2006synchronization}, we computed the LE on more than 1,000 independent noise realizations for each position of the wall. The mean LE values and their standard deviations are plotted in Fig.~\ref{fig:lyap}, with sky-blue points representing the mean and error bars indicating the standard deviation. In the above-mentioned ranges of $x_w$, the mean LE remains positive and exceeds the standard deviation, indicating robust instability across noise realisations.

To confirm the presence of chaos, we applied the 0–1 test, which is effective even in stochastic systems \cite{gottwald2004new}. As shown in Fig.~\ref{fig:01}, the test yields values close to 1 within the same parameter regime, consistent with chaotic dynamics.

 In Figs.~\ref{fig:bif}, \ref{fig:lyap}, and \ref{fig:01}, these wall positions are marked by green stars, purple diamonds, and red triangles for visual reference to trajectories A, B, and C shown in Fig.~\ref{fig:traj_all_walls}.

\section{Concluding Thoughts and Future Directions\label{conclusion}}

In this paper, we have demonstrated that the quantum analog of an impact oscillator exhibits a rich array of dynamical behaviors. 
\begin{itemize}
    \item A hard impact oscillator without dissipation or forcing displays a quasiperiodic behavior;
    \item A hard impact oscillator without dissipation but with sinusoidal forcing can exhibit strange non-chaotic behavior;
    \item A soft impact oscillator with dissipation and forcing can exhibit chaotic dynamics.
\end{itemize}

This work provides the first numerical evidence for strange non-chaotic dynamics in a quantum system. Similar analyses of other forced quantum systems such as the pendulum, Duffing oscillator, and Kapitza pendulum do not show signs of strange nonchaotic behavior.

The adaptation of tools from nonlinear dynamics has profoundly enriched our understanding of quantum chaos. By moving beyond a one-to-one mapping between the classical dynamics and the quantum level-spacing statistics and embracing creative new measures and representations, we have been able to uncover the subtle quantum signatures of classical chaos.

This work has highlighted the multifaceted nature of quantum chaos, revealing that it cannot be captured by any single diagnostic measure. Instead, we have seen how a web of interconnected diagnostics---from spectral statistics and OTOCs to 0-1 test---provide a comprehensive picture of chaotic quantum behavior. 


For the nonlinear dynamics community, the study of quantum chaos offers a fertile and exciting frontier. The tools and concepts that we are familiar with are not just applicable, but essential for unraveling the mysteries of the quantum world. As we continue to push the boundaries of both fields, we can expect many more surprising and profound discoveries to emerge from this vibrant interplay. The recent developments in quantum technologies, from quantum computers to quantum simulators, provide new experimental platforms for testing theoretical predictions and exploring quantum chaos in unprecedented detail.

\begin{table}
    \centering
    \begin{tabular}{|p{0.45\textwidth}|p{0.45\textwidth}|}
    \hline 
    \textbf{Classical} & \textbf{Quantum} \\
    \hline
    Largest Lyapunov exponent & Exponent of the initial growth of the out-of-time-order correlator \\
    \hline
    Frequency spectrum of the time series & Eigenvalue spectrum of the (Floquet) Hamiltonian operator for static (time-periodic) Hamiltonians 
    \\
    \hline
    Bifurcation diagram is plotted using a Poincaré section or stroboscopic samples of the dynamical variables & Bifurcation diagram can be plotted by averaging over stroboscopic samples of the probability density \\
    \hline
    \end{tabular}
    \caption{A summary of key tools from nonlinear dynamics and their quantum (quasi)-equivalents.}
    \label{tab:tool_comparison}
\end{table}

\section*{Appendix~A: Commutator-free Exponential Time-propagators (CFETs)}

In this case, we have to solve the Schr\"odinger equation 
\begin{equation}
    i\hbar\frac{\partial}{\partial t} \Psi(t) = H(t)\Psi(t), \label{schr2}
\end{equation}
where $H(t)$ is a time-dependent Hamiltonian operator.
Equation (\ref{schr2}) is a special case of a general linear differential equation of the form,
\begin{equation}
    \frac{\partial}{\partial t} x(t) = A(t)x(t), \label{diff.eq}
\end{equation}
In the context of the Schr\"odinger equation, $A(t)$ is defined as $A(t) = -i\frac{H(t)}{\hbar}$. Traditional numerical techniques often rely on methods like the Magnus expansion, which provides a systematic way to approximate the time-evolution operator.
The solution to (\ref{diff.eq}) is written as follows:
 \begin{equation}
     x(\delta t)  = U(\delta t, 0)x(0) 
 \end{equation}
The propagator $U(\delta t,0)$ is determined using the Magnus expansion that involves commutators, which can be computationally expensive to evaluate, especially for higher-order terms. To overcome that, we have used commutator-free exponential propagators (CFETs) \cite{ALVERMANN20115930}.
CFETs do not use commutators that appear in the Magnus series; the propagators are approximated using the Baker-Campbell-Hausdorff formula (BCH).
The 4th-order optimized commutator-free exponential time propagator (CFET)\cite{Alvermann_2012} is given by
\begin{align}
U(\delta t) &= \exp\left[ \delta t \left( g_1 A^{(1)} + g_2 A^{(2)} + g_3 A^{(3)} \right) \right] \notag \\
&\quad \times \exp\left[ \delta t \left( g_4 A^{(1)} + g_5 A^{(2)} + g_4 A^{(3)} \right) \right] \notag \\
&\quad \times \exp\left[ \delta t \left( g_3 A^{(1)} + g_2 A^{(2)} + g_1 A^{(3)} \right) \right], 
\end{align}
where
\begin{equation}
A^{(1)} = A[x_1 \:\delta t], \quad A^{(2)} = A[x_2 \:\delta t], \quad A^{(3)} = A[x_3 \: \delta t]
\end{equation}
and
\begin{equation}
x_1 = \frac{1}{2} - \sqrt{\frac{3}{20}}, \quad x_2 = \frac{1}{2}, \quad x_3 = \frac{1}{2} + \sqrt{\frac{3}{20}} 
\end{equation}

\begin{equation}
\begin{aligned}
g_1 &= \frac{37}{240} - \frac{10}{87} \sqrt{\frac{5}{3}}, \\
g_2 &= -\frac{1}{30}, \\
g_3 &= \frac{37}{240} + \frac{10}{87} \sqrt{\frac{5}{3}}, \\
g_4 &= -\frac{11}{360}, \\
g_5 &= \frac{23}{45}.
\end{aligned} 
\end{equation}

\section*{Appendix~B: The semiclassical quantum Langevin equation for modeling the system with dissipation}

The equations of motion in this formalism are given by 
\begin{equation}
 \begin{aligned}
 \dot{X} &= P, \\
 \dot{P} &= -V'(X,t) + f(t) + z + Q(t), \\
 \dot{z} &= -\Gamma \frac{P}{\tau_c} - \frac{z}{\tau_c}.
 \end{aligned}
 \label{eqn: cQLE}
 \end{equation}
Here, $X(t) = \langle \hat{x}(t) \rangle$ and $P(t) = \langle \hat{p}(t) \rangle$ denote the quantum mechanical expectation values of the position and momentum operators. The term $V'(X,t)$ represents the first derivative of the potential with respect to the position.

Dissipation is captured by the auxiliary variable $z(t)$, which encodes non-Markovian memory effects arising from the system-bath interaction. The parameters $\Gamma$ and $\tau_c$ represent the dissipation strength and the bath correlation time, respectively. The latter determines the timescale over which the bath retains memory of its past interactions with the system: a shorter $\tau_c$ corresponds to a faster decay of correlations.

The stochastic force $f(t)$ is modeled as the sum of exponentially correlated colored noise originating from the thermal reservoir.
\begin{equation}
f(t) = \sum_{i=1}^n \eta_i(t), \quad \dot{\eta}_i = -\frac{\eta_i}{\tau_i} + \frac{1}{\tau_i} \xi_i(t),
\end{equation}
where each $\eta_i$ has a finite correlation time $\tau_i$, and $\xi_i(t)$ is a Gaussian white noise of zero mean, where these noise properties depend on the temperature ($kT$) of the thermal reservoir.

The term $Q(t)$ accounts for purely quantum effects that arise when the potential of the system is not a simple linear or quadratic function. In such cases, quantum particles behave differently from classical ones because they experience additional corrections due to fluctuations around their average position.
Mathematically, this correction is defined as follows.
\begin{eqnarray}
Q(t) &=& V'(X,t) - \langle V'(\hat{x},t) \rangle \nonumber \\ &=& -\sum_{n \geq 2} \frac{1}{n!} V^{(n+1)}(X) \langle \delta \hat{x}^n(t) \rangle,
\label{eqn:Q}
\end{eqnarray}
where $V'(X,t)$ is the force from the potential evaluated at the mean position $X(t)$, and $\langle V'(\hat{x},t) \rangle$ is the average force felt by the quantum particle. The difference between them gives $Q(t)$, which depends on the shape of the potential by the term $V^{(n+1)}$, the $(n+1)$th derivative of the potential with respect to $x$ and the amount of fluctuation around the mean position by $\langle \delta\hat {x}^n(t) \rangle$, the quantum mechanical average of the $n$th order correction to the position operator. The evolution of $\langle \delta\hat {x}^n(t) \rangle$ depends on $\hbar$ according to the uncertainty principle. For further details, see \cite{banerjee2003numerical,barik2003numerical,mukherjee2025dynamical}. 
\section*{Acknowledgements}
S Banerjee and A Acharya acknowledge financial support
from J. C. Bose Grant of the Anusandhan National Research Foundation, Govt. of India, No.
JBR/2020/000049. T Mukherjee acknowledges financial support from  UGC, Govt. of India.

\end{document}